\documentstyle[twoside,fleqn,espcrc2,psfig]{article}

\def\dash{\hbox{--}}
\def\degree{{$^\circ$}}

\def\puncspace{\ifmmode\,\else{\ifcat.\C{\if.\C\else%
\if,\C\else\if?\C\else\if:\C\else\if;\C\else\if-\C\else%
\if)\C\else\if/\C\else\if]\C\else\if'\C%
\else\space\fi\fi\fi\fi\fi\fi\fi\fi\fi\fi}%
\else\if\empty\C\else\if\space\C\else\space\fi\fi\fi}\fi}%
\def\SP{\let\\=\empty\futurelet\C\puncspace}

\def\ee#1{\ifmmode {} \times 10^{#1} \else ${} \times 10^{#1}$\fi}
\def\sub#1{\ifmmode _{#1} \else $_{#1}$\fi}
\def\sup#1{\ifmmode ^{#1} \else $^{#1}$\fi}

\def\dash{\hbox{--}}
\def\about{\ifmmode \sim \else {$\sim\,$}\fi}
\def\lta{\ifmmode {\,\mathbin{\lower 3pt\hbox   
    {$\,\rlap{\raise 5pt\hbox{$\char'074$}}\mathchar"7218\,$}}}
    \else {${\mathbin{\lower 3pt\hbox
    {$\rlap{\raise 5pt\hbox{$\char'074$}}\mathchar"7218\,$}}}
    $}\fi}
\def\gta{\ifmmode {\mathbin{\lower 3pt\hbox   
    {$\,\rlap{\raise 5pt\hbox{$\char'076$}}\mathchar"7218\,$}}}
    \else {${\mathbin{\lower 3pt\hbox
    {$\rlap{\raise 5pt\hbox{$\char'076$}}\mathchar"7218\,$}}}
    $}\fi}

 \mathcode`*="002A   

\def\degree{{\ifmmode ^\circ \else $^\circ$\fi}}

\def\sco#1{\leavevmode\hbox{Sco~X-#1}\SP}

\def\exosat{\leavevmode{\it EXOSAT}\SP}
\def\ginga{\leavevmode{\it Ginga\/}\SP}

\def\rxte{\leavevmode{\it RXTE}\SP}

\def\mdot{{\ifmmode \dot M \else {$\dot M$}\fi}}
\def\mdote{{\ifmmode \dot M_E \else {$\dot M_E$}\fi}}
\def\mdoti{{\ifmmode \dot M_i \else {$\dot M_i$}\fi}}
\def\msun{{\ifmmode M_\odot \else {$M_{\odot}$}\fi}}

\title{Rapid X-Ray Variability of Neutron
Stars in Low-Mass Binary Systems}

\author{Frederick K. Lamb\address{University of Illinois
        at Urbana-Champaign,
        Department of Physics and Department of Astronomy,\\
        1110 W. Green St.,
        Urbana, IL  61801,
        USA}
        M. Coleman Miller\address{University of Chicago,
        Department of Astronomy and Astrophysics,
        5640 S. Ellis Avenue,\\
        Chicago, IL  60637,
        USA}
        and
        Dimitrios Psaltis\address{Harvard-Smithsonian
        Center for Astrophysics,
        60 Garden St.,
        Cambridge, MA 02138,
        USA}}

\begin{document}

 \begin{abstract}
 The dramatic discovery with the {\em Rossi
X-Ray Timing Explorer\/} satellite of
remarkably coherent $\sim$300--1200~Hz
oscillations in the X-ray brightness of
some sixteen neutron stars in low-mass
binary systems has spurred theoretical
modeling of these oscillations and
investigation of their implications for the
neutron stars and accretion flows in these
systems. High-frequency oscillations are
observed both during thermonuclear X-ray
bursts and during intervals of
accretion-powered emission and appear to be
a characteristic feature of disk-accreting
neutron stars with weak magnetic fields. In
this review we focus on the high-frequency
quasi-periodic oscillations (QPOs) seen in
the accretion-powered emission. We first
summarize the key properties of these
kilohertz QPOs and then describe briefly
the models that have been proposed to
explain them. The existing evidence
strongly favors beat-frequency models. We
mention several of the difficulties
encountered in applying the magnetospheric
beat-frequency model to the kilohertz QPOs.
The most fully developed and successful
model is the sonic-point beat-frequency
model. We describe the work on this model
in some detail. We then discuss
observations that could help to distinguish
between models. We conclude by noting some
of the ways in which study of the kilohertz
QPOs may advance our understanding of dense
matter and strong gravitational fields.
 \end{abstract}

\maketitle

\section{INTRODUCTION}

It has long been expected (see, e.g.,
\cite{ELP86,LP79}) that important
information about the intrinsic properties
of neutron stars and stellar mass black
holes, as well as about the physics of
accretion onto them, could be extracted
from their X-ray variability at frequencies
comparable to the \hbox{$\sim$1--10~kHz}
dynamical frequencies near them. The {\em
Rossi X-ray Timing Explorer\/} (\rxte\/) was
specifically designed \cite{BS89,Sw95} to
have the large area, microsecond time
resolution, and high telemetry rate needed
to probe this high-frequency regime.

 \begin{figure}[t!] 
 \vglue0.2truecm
 \centerline{\hskip0.25truecm
\psfig{file=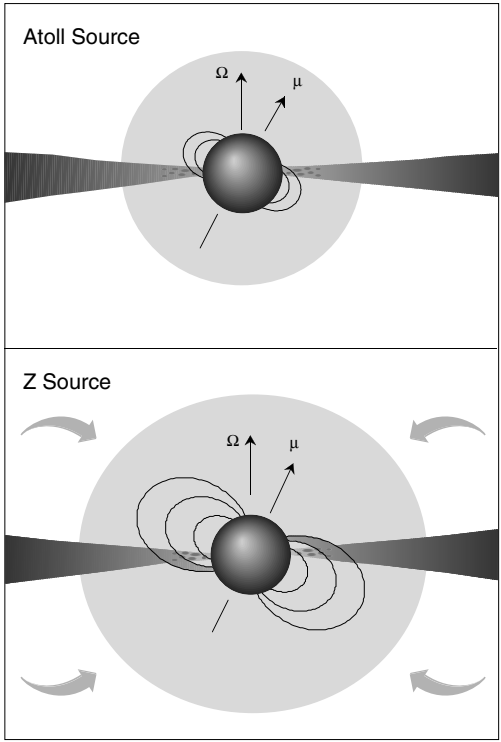,angle=0,width=6.6truecm}}
 \vglue-1.0truecm
 \caption{\label{fig1}
 Side view of an atoll source (top) and a Z
source (bottom). The dark shading indicates
the accretion disk. The light shading
indicates the hot gas in the magnetosphere
and the central corona that surrounds the
neutron stars. The arrows in the lower
panel indicate the cooler, approximately
radial inflow that is thought to be present
outside the central coronae of the Z
sources. From [28].
 }
 \label{nsLMXBs}
 \end{figure}

The value of access to these high
frequencies was dramatically confirmed
when, less than two months after the launch
of \rxte, remarkably coherent $\sim
1000$~Hz oscillations were discovered (see
\cite{vdK97b}) in the X-ray brightness of
two neutron stars in low-mass X-ray binaries
(LMXBs). These are the fastest astrophysical
oscillations ever discovered. It is thought
that they are being generated at or near
the surfaces of these neutron stars.

To date, brightness oscillations with
frequencies ranging from $\sim$300~Hz to
more than 1200~Hz have been discovered in
some sixteen neutron stars in LMXBs.
High-frequency oscillations are observed
both during thermonuclear X-ray bursts
\cite{Stroh96c,Stroh97,Swank98} and during
intervals of accretion-powered emission
\cite{vdK97b,vdK98}. They appear to be a
characteristic feature of disk-accreting
neutron stars with weak magnetic fields.

In this review we focus on the
high-frequency quasi-periodic oscillations
(QPOs) seen in the accretion-powered
emission. The weak-field neutron stars in
which these {\em kilohertz QPOs\/} have
been discovered were studied extensively
with \exosat\ and \ginga, and a detailed
physical picture was developed based on
their 2--20~keV X-ray spectra and 1--100~Hz
X-ray variability (see
\cite{GL92,HK89,L89,L91}). In this picture,
LMXBs with weak-field neutron stars fall
into two classes, called the ``Z'' and
``atoll'' sources after the shapes of the
paths they trace, over time, in X-ray
color-color diagrams \cite{HK89}. Modeling
\cite{PL97,PL98,PLM95} of the X-ray spectra
of these sources indicates that both types
are surrounded by a central corona with
a scattering optical depth $\sim$3--10, and
that the Z sources also have a cooler
radial inflow (see Fig.~\ref{nsLMXBs}).

The six known ``Z'' sources have mass
accretion rates comparable to the Eddington
critical rate $\mdote$ and inferred surface
magnetic fields
\about$10^{9}$--$10^{10}\,$G. They display
two different types of lower-frequency
QPOs. The $\sim$15--50~Hz
``horizontal-branch oscillations'' (HBOs)
\cite{vdK85} have frequencies that increase
steeply with accretion rate and are thought
to be produced by the magnetospheric
beat-frequency mechanism
\cite{AS85,GL92,L85}. The $\sim$4--8~Hz
``normal/flaring branch oscillations''
(N/FBOs) \cite{MP86} are thought to be
caused by radiation-hydrodynamic
oscillations in the radial inflow
\cite{FLM89,L89,L91}.

The $\sim$15 known ``atoll'' sources are
both less luminous and more weakly
magnetic than the Z sources, with
accretion rates $\sim$1--10\% of $\mdote$
and magnetic fields $\sim 10^7 \dash\,
5\ee9\,$G. No QPOs with frequencies $\lta
100\,$Hz have so far been detected in any
of the atoll sources, with the exception of
Cir~X-1, which has a $\sim$1--30~Hz QPO
with a frequency that varies with its
brightness (see \cite{Bradt98}).

Considerable theoretical effort has been
devoted to understanding the mechanisms
that produce the kilohertz QPOs. In this
review we first summarize the key
properties of these QPOs and then describe
briefly the models that have been proposed
to explain them. The kilohertz QPOs
commonly occur in pairs, and the existing
evidence strongly favors beat-frequency
models. We discuss application of the
magnetospheric beat-frequency model to the
kilohertz QPOs. The most fully developed
and successful model is the so-called
sonic-point beat-frequency model, in which
the higher frequency in a QPO pair is the
orbital frequency of gas at the inner edge
of the Keplerian disk flow and the lower
frequency is the difference between this
frequency and the spin frequency of the
neutron star. We outline the sonic-point
model and describe some of the calculations
that have been carried out to explore it.
We conclude by discussing observations that
could help to distinguish between models
and noting some of the ways in which the
study of kilohertz QPOs may advance our
understanding of neutron stars, dense
matter, and strong gravitational fields.

\section{CHARACTERISTICS OF KHZ QPOs}

The kilohertz QPOs observed in the
neutron-star LMXBs have strikingly similar
properties \cite{vdK97b,vdK98}. These
properties include:

{\em High frequencies}.---The frequencies
of the kilohertz QPOs range from
$\sim$300~Hz to $\sim$1200~Hz.

{\em High maximum amplitudes}.---The
strongest kilohertz QPOs have fractional rms
amplitudes as high as $\sim$15\% of the
total 2--60~keV countrate. The amplitudes of
the kilohertz QPOs are much larger in the
atoll sources than in the Z sources, where
they are typically $\lta 1$\%.

{\em High coherence}.---QPOs with quality
factors $Q\equiv \nu_{\rm QPO}/
\Delta\nu_{\rm QPO} \sim 50$--100 are
common; in some sources, the kilohertz QPOs
have $Q$ values as high as 200.

{\em Frequency-\mdot correlation}.---In the
Z sources, the frequencies of the kilohertz
QPOs increase steeply with \mdot. In many
atoll sources, the frequencies of the
kilohertz QPOs increase steeply with
countrate (in at least some, the
frequency-countrate relation shifted
significantly between observations made just
a few days apart \cite{Ford97b,Mend97b}).
Although the magnetic field strengths and
accretion rates of the atoll and Z sources
are thought to be quite different, their
frequency-countrate and frequency-accretion
rate correlations are very similar.

{\em Similar frequency ranges}.---The
frequency ranges of the kilohertz QPOs are
similar in the Z and atoll sources, even
though these two types of sources are
thought to have very different magnetic
field strengths and accretion rates.

{\em QPO pairs}.---Kilohertz QPOs commonly
occur in pairs and, except in Sco~X-1
\cite{vdK97b} (and possibly in 4U~1608--52
\cite{Mend97b}), maintain a constant
frequency separation $\Delta\nu$ as they
shift up and down in frequency by hundreds
of Hertz.

{\em Relation to spin frequency}.---X-ray
brightness oscillations have been observed
during X-ray bursts in four sources in
which kilohertz QPO pairs have been detected
(see \cite{MLP97,vdK97b}). The frequency
$\nu_{\rm burst}$ of the burst oscillation
appears to be the stellar spin frequency
$\nu_{\rm spin}$ or twice $\nu_{\rm spin}$
\cite{Stroh97,Swank98}. In these four
sources, the separation $\Delta\nu$ between
the kilohertz QPOs appears to be $\nu_{\rm
burst}$ or half $\nu_{\rm burst}$.

{\em Photon-energy dependence}.---The
relative amplitudes of many kilohertz QPOs
increase steeply with photon energy in the
2--15~keV energy band.

In order to be successful, a model of the
kilohertz QPOs must explain these
characteristics and be consistent with the
data obtained previously on the 2--20~keV
X-ray spectra and 1--100~Hz X-ray
variability of the Z and atoll sources. We
now review the proposed models, comparing
them with these requirements.

\section{KILOHERTZ QPO MODELS}

\subsection{Photon bubble oscillations}

Photon bubble oscillations (PBOs) were first
proposed as a model for X-ray brightness
variations in accretion-powered pulsars (see
\cite{Klein96a} and references therein),
where the strong stellar magnetic field
funnels the flow onto the magnetic polar
caps, producing a mass flux onto the caps
that is locally super-Eddington. Under these
conditions, photon ``bubbles'' form in the
accretion column above the caps, rise, and
burst, causing brightness variations. In the
initial simulations \cite{Klein96a}, which
considered an accretion rate of
$0.2\,\mdote$ and a surface dipole magnetic
field of $3\ee{12}$~G, brightness
fluctuations occurred over a narrow range of
frequencies at a few thousand Hertz.

Klein et~al.\ \cite{Klein96b} suggested that
the same mechanism is responsible for the
kilohertz QPO pairs, and predicted that
additional QPOs would be found at higher
frequencies. Observations of \sco1
performed to check this \cite{vdK97a} showed
no evidence of higher-frequency
oscillations down to fairly small
amplitudes. However, if there is scattering
material around the neutron star
magnetosphere, as inferred from \exosat\
and \ginga\ observations (see Fig.~1), the
observed amplitudes of PBOs with
frequencies above a kilohertz would be
significantly reduced. In more recent
simulations carried out for a magnetic
field of $5\ee9$~G, one or two QPOs with
rms amplitudes up to 3\% occurred, with
frequencies that increased with increasing
accretion rate, accompanied by weaker
higher-frequency QPOs \cite{Klein97}. The
relevant three-dimensional
radiation-hydrodynamics is very
complicated, making it hard to achieve
analytic insights (but see \cite{Arons92})
and difficult to survey the full parameter
space numerically.

The locally super-Eddington mass fluxes
required for photon bubbles to form are not
expected in the atoll sources, which are
thought to have weak magnetic fields and
total accretion rates as low as $\sim
0.003\,\mdote$, but suitable conditions may
exist in the Z sources. If the Z-source
kilohertz QPOs are PBOs, then the
atoll-source kilohertz QPOs, which appear
very similar (see \S2), would have to be
generated by a different mechanism. It
remains to be seen whether PBOs can explain
key characteristics of the kilohertz QPO
pairs, such as their constant frequency
separation and the steep increase in their
relative amplitude with increasing photon
energy.

\subsection{Disk oscillations}

Titarchuk \& Muslimov \cite{TM1997} have
proposed that the frequencies of the burst
oscillation and the two kilohertz QPOs seen
in many sources (see \S2) are the
frequencies of three specific oscillation
modes of the accretion disk. In this model,
as in all models in which the kilohertz
QPOs are produced by disk oscillations, it
is difficult to understand how QPOs with
amplitudes as large as 15\% of the total
countrate can be generated, because
typically $\lta 20$\% of the accretion
luminosity is released in the accretion
disk. In addition, one must explain why
only three of the large number of disk
oscillation modes produce large-amplitude
brightness oscillations.

No proposal was made by Titarchuk \&
Muslimov for how the modes they discussed
are excited, why only the modes at a
particular radius are excited, or how they
produce large-amplitude brightness
oscillations. Moreover, the expression for
mode frequencies that they used generally
predicts that the {\em ratio\/} of the
frequencies of two different modes will
remain constant as the frequencies change,
whereas it is the {\em separation\/}
between the kilohertz QPOs in a pair that
is nearly constant. For the frequency
separation to be constant in this model,
one must appeal either to a coincidence or
to some physical principle that has not yet
been elucidated. A more serious difficulty
is that the method \cite{Unno79} used by
Titarchuk \& Muslimov to compute the
frequencies of oscillation modes in the
centrifugally supported, strongly sheared
accretion disk flow is a perturbation
method designed for computing the small
changes in the frequencies of the nonradial
oscillation modes of a spherical,
gas-pressure-supported star caused by slow,
uniform rotation.

 \begin{figure*}[t] 
 \vglue-1truecm
\hbox{
\hglue0.1truecm
{\psfig{file=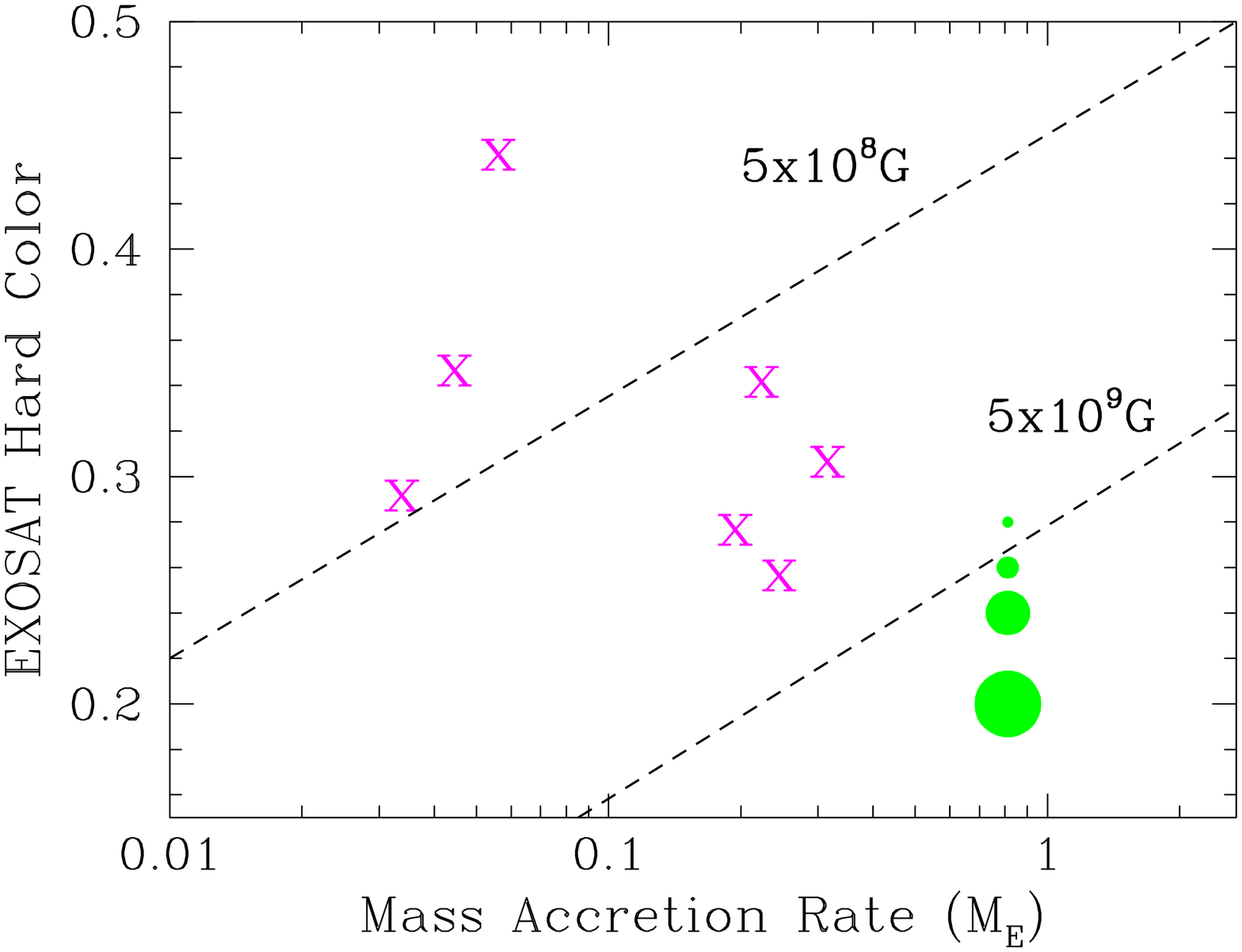,angle=270,height=2.7truein,width=3.6truein}}
\hglue-1.5truecm
{\psfig{file=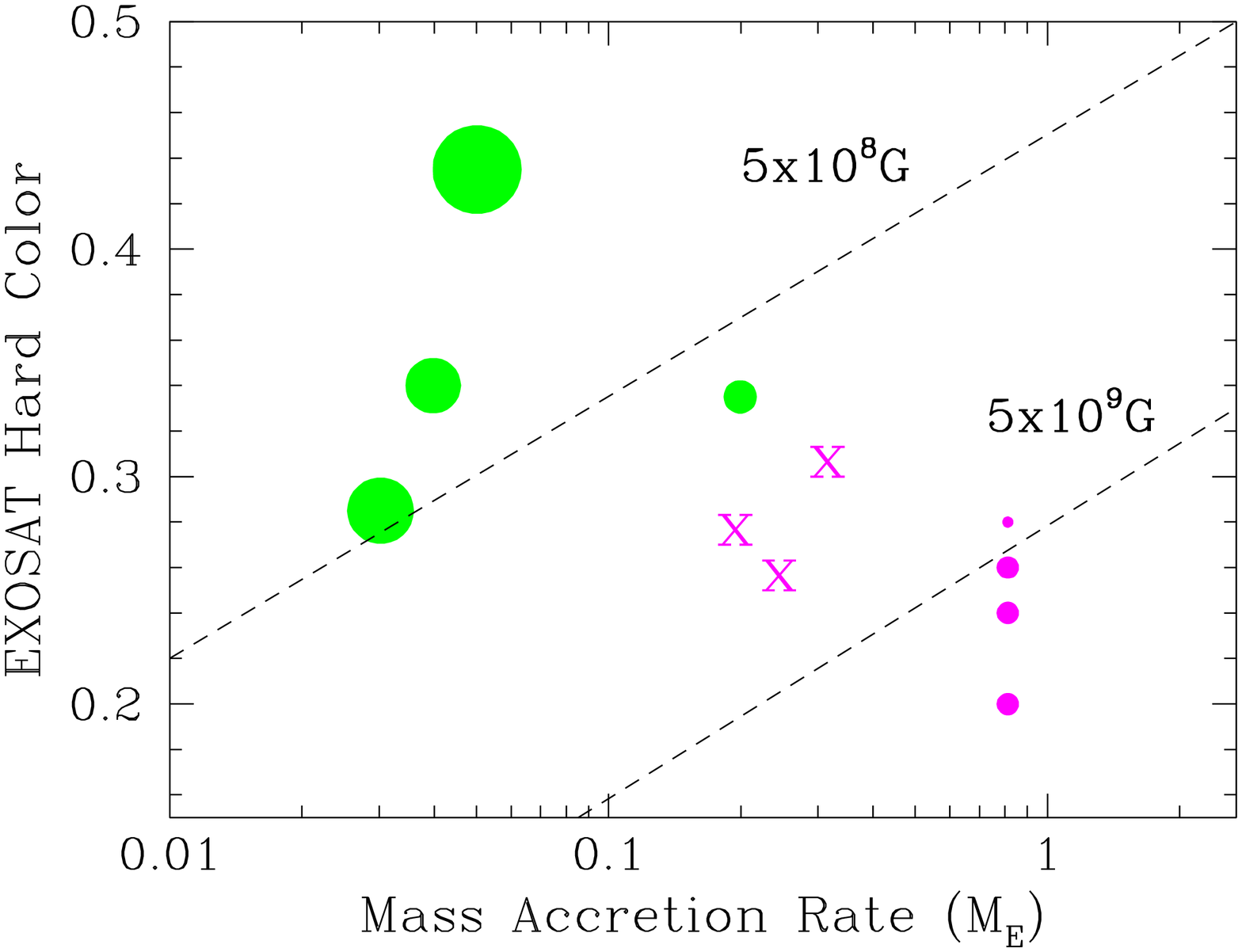,angle=270,height=2.7truein,width=3.6truein}}}
 \vglue-1.25truecm
 \caption{
 Correlation of the observed amplitudes of
HBOs (left panel) and kilohertz QPOs (right
panel) with the strength of the neutron
star magnetic field (dashed lines)
inferred from detailed radiation transport
calculations of the X-ray spectrum
[30--32]. The areas of the shaded circles
are proportional to the measured amplitudes
of the QPOs. The X's indicate that no QPO
has been detected.
 The three sources in the upper left are,
from top to bottom, 4U~1728$-$34,
4U~1820$-$30, and 4U~1636$-$536. The four
sources in the center are 4U~1735$-$444,
GX~3$+$1, GX~9$+$9, and GX~13$+$1. The four
sources in the lower right are the Z sources
Sco~X-1, GX~17$+$2, GX~5$-$1, and Cyg~X-2.
From [31].
 } \label{qpoAmps} \end{figure*}

\subsection{Magnetospheric Keplerian
frequency and beat-frequency oscillations}

The magnetospheric beat-frequency model was
developed \cite{AS85,L85} to explain the
single, $\sim$15--60~Hz ``horizontal branch 
oscillation'' (HBO) observed in the Z
sources (see \S1). In this model, the
frequency of the HBO is the difference
between the Keplerian orbital frequency
$\nu_{\rm Km}$ at the main radius where the
stellar magnetic field picks up and
channels gas from the accretion disk onto
the magnetic polar regions (the ``main gas
pick-up radius'') and the stellar spin
frequency $\nu_{\rm spin}$. In applying the
magnetospheric beat-frequency idea to the
kilohertz QPO pairs, Strohmayer et~al.\
\cite{Stroh96c} interpreted the frequency
of the higher-frequency QPO in a pair as
$\nu_{\rm Km}$ and the frequency of the
lower-frequency QPO as $\nu_{\rm Km} -
\nu_{\rm spin}$.

An attractive feature of the magnetospheric
beat-frequency interpretation of the
kilohertz QPO pairs is that it explains
naturally why the frequency separation
between the QPOs in a pair is nearly
constant in most sources and equal to the
burst oscillation frequency or half this
frequency. However, there are many serious
difficulties with this interpretation
\cite{MLP97}. For example, no mechanism has
been found that will generate a relatively
coherent QPO at the orbital frequency at
the main gas pick-up radius. The high
observed coherence of the kilohertz QPOs is
also a serious difficulty for the
magnetospheric beat-frequency
interpretation. HBOs and kilohertz QPO
pairs occur simultaneously in at least five
of the six known Z sources (see
\cite{vdK98}), so the magnetospheric
beat-frequency model cannot explain both
types of QPO. The fact (see
Fig.~\ref{qpoAmps}) that HBOs are seen only
in the Z sources, which are thought to have
magnetic fields $\sim 10^{9} \dash
10^{10}\,$G, and have higher amplitudes in
the Z sources that have stronger magnetic
fields, whereas the kilohertz QPOs are much
stronger in the atoll sources, which are
thought to have magnetic fields $\sim 10^7
\dash 5\ee9$~G, is strong evidence that the
HBOs are magnetospheric and that the
kilohertz QPOs are not.

\subsection{Sonic-point Keplerian
frequency and beat-frequency oscillations}

The sonic-point model \cite{MLP97} was
developed specifically to explain the
kilohertz QPO pairs. In this model, some
accreting gas spirals inward in nearly
circular Keplerian orbits until it is close
to the neutron star, where radiation forces
or general relativistic effects cause a
sudden increase in the inward radial
velocity, which becomes supersonic within a
small radial distance. The sharp increase in
the radial velocity is usually caused by the
drag exerted on the accreting gas by
radiation from the star, but may instead be
caused by general relativistic corrections
to Newtonian gravity if the gas in the
Keplerian disk flow reaches the innermost
stable circular orbit (ISCO) without being
significantly affected by radiation. For
convenience, the radius at which the radial
velocity increases is referred to as the
``sonic point'', even though the sonic
point itself is not especially significant
in this model.

In the sonic-point model, the frequency of
the higher-frequency QPO in a pair is the
orbital frequency $\nu_{\rm Ks}$ near the
sonic radius. Density fluctuations at this
radius are mapped onto the stellar surface,
producing an X-ray beam that rotates around
the star with frequency $\nu_{\rm Ks}$. A
distant observer therefore sees
quasi-periodic X-ray spectral and
brightness oscillations with frequency
$\nu_{\rm Ks}$.

The frequency of the lower-frequency QPO in
a pair is the difference between this
orbital frequency and the stellar spin
frequency. The lower-frequency QPO is
generated by weak X-ray beams produced by
funneling of part of the accretion flow
near the star by the star's weak magnetic
field. These beams rotate {\em with the
star\/} and modulate the radiation drag
acting on the gas at the sonic radius,
modulating the inward mass flux and the
luminosity at the sonic-point beat
frequency $\nu_{\rm Bs}$ ($\nu_{\rm
Ks}-\nu_{\rm spin}$ or $\nu_{\rm
Ks}-2\nu_{\rm spin}$).

 \begin{figure*}[t!] 
 \vglue-2.2truecm
 \centerline{\hskip0.5truecm
\psfig{file=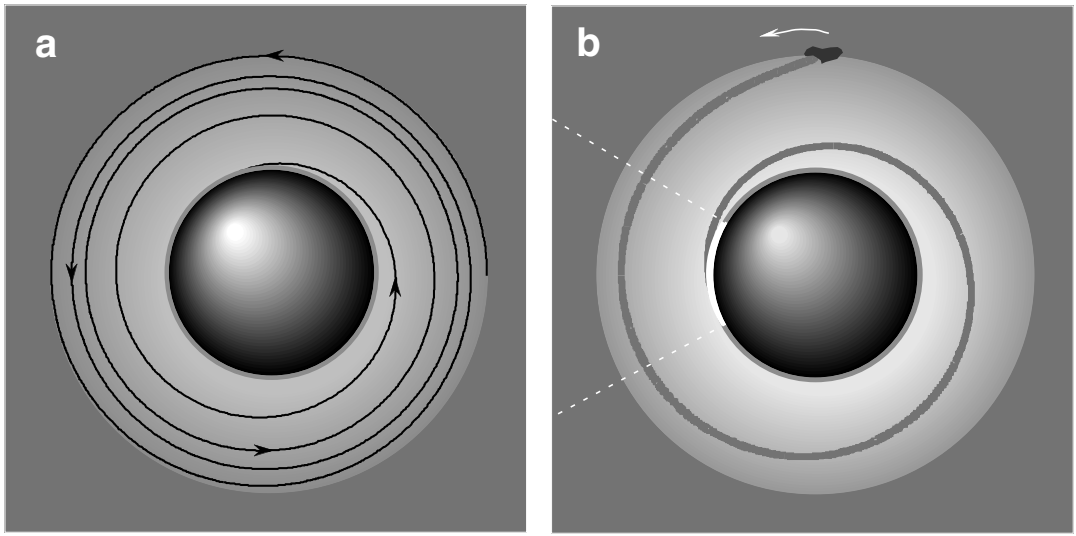,angle=90,height=13.0truecm}}
 \vglue-4.2truecm
 \caption{View of the neutron star and disk
along the rotation axis of the disk,
which is rotating counterclockwise in this
view.
  (a)~The spiral trajectory
followed by a single element of gas as it
falls supersonically from the sonic radius
to the stellar surface.
  (b)~The spiral pattern of higher gas
density formed by gas streaming inward
along spiral trajectories with the shape
shown in (a), from a clump orbiting
near the sonic radius.
 The surface density of the disk flow is
much smaller inside the sonic radius
(lighter shaded region) than outside
(darker shaded region), because of the
sharp increase in the inward radial
velocity at the sonic radius.
 The spiral trajectory and density pattern
are from fully general relativistic
numerical computations of the gas dynamics
and radiation transport in the inner disk
[28].
 } \label{spirals} \end{figure*}

The sonic-point model is consistent with
the accretion rates, stellar magnetic
fields, and scattering optical depths
inferred previously from \exosat\ and
\ginga\ observations of the atoll and Z
sources and accounts for the main features
of the kilohertz QPOs (see \cite{MLP97}),
including their high and variable
frequencies, their high amplitudes and
coherences, and the common occurrence of
kilohertz QPOs in pairs. Like the
magnetospheric beat-frequency
interpretation, the sonic-point
beat-frequency model explains naturally why
the frequency separation between the
frequencies of a kilohertz QPO pair is
nearly constant and equal to the burst
oscillation frequency or half this
frequency.

The sonic-point model also explains why the
frequency ranges of the kilohertz QPOs are
similar in sources with very different
accretion rates and magnetic fields (see
\S4). Finally, it accounts (see
\cite{MLP97}) for the observed increase in
QPO amplitude with increasing photon energy
and the observed anti-correlation between
kilohertz QPO amplitude and stellar
magnetic field strength (see
Fig.~\ref{qpoAmps}).

There are several unresolved questions
related to the sonic-point model. Perhaps
the most important is what fraction of the
gas in the accretion disk does not couple
strongly to the stellar magnetic field and
therefore spirals inward in nearly circular
orbits until it is close to the neutron
star. This question is particularly
important for the Z sources, some of which
have inferred magnetic fields as high as
$10^{10}\,$G. Although it has been realized
for more than two decades (see, e.g.,
\cite{AL76a,EL76,GL79a,L75b})
that instabilities will allow some fraction
of the accreting gas to penetrate deep
inside the magnetosphere, it has not been
possible to calculate this fraction
reliably or to determine how much of this
gas remains in nearly circular orbits. The
kilohertz QPOs are strong evidence that at
least some gas is orbiting close to the
star, because the orbital frequencies at
the radii where gas is expected to begin
coupling to the stellar magnetic field are
$\sim$5--10 times lower than the
frequencies of the kilohertz QPOs.

The sonic-point model is currently the most
fully developed and successful explanation
for the kilohertz QPOs. In the next section
we describe some of the calculations that
have been carried out to explore it.

 \begin{figure*}[t] 
 \vglue-1.5truecm
\hbox{\hglue-0.2truecm
{\psfig{file=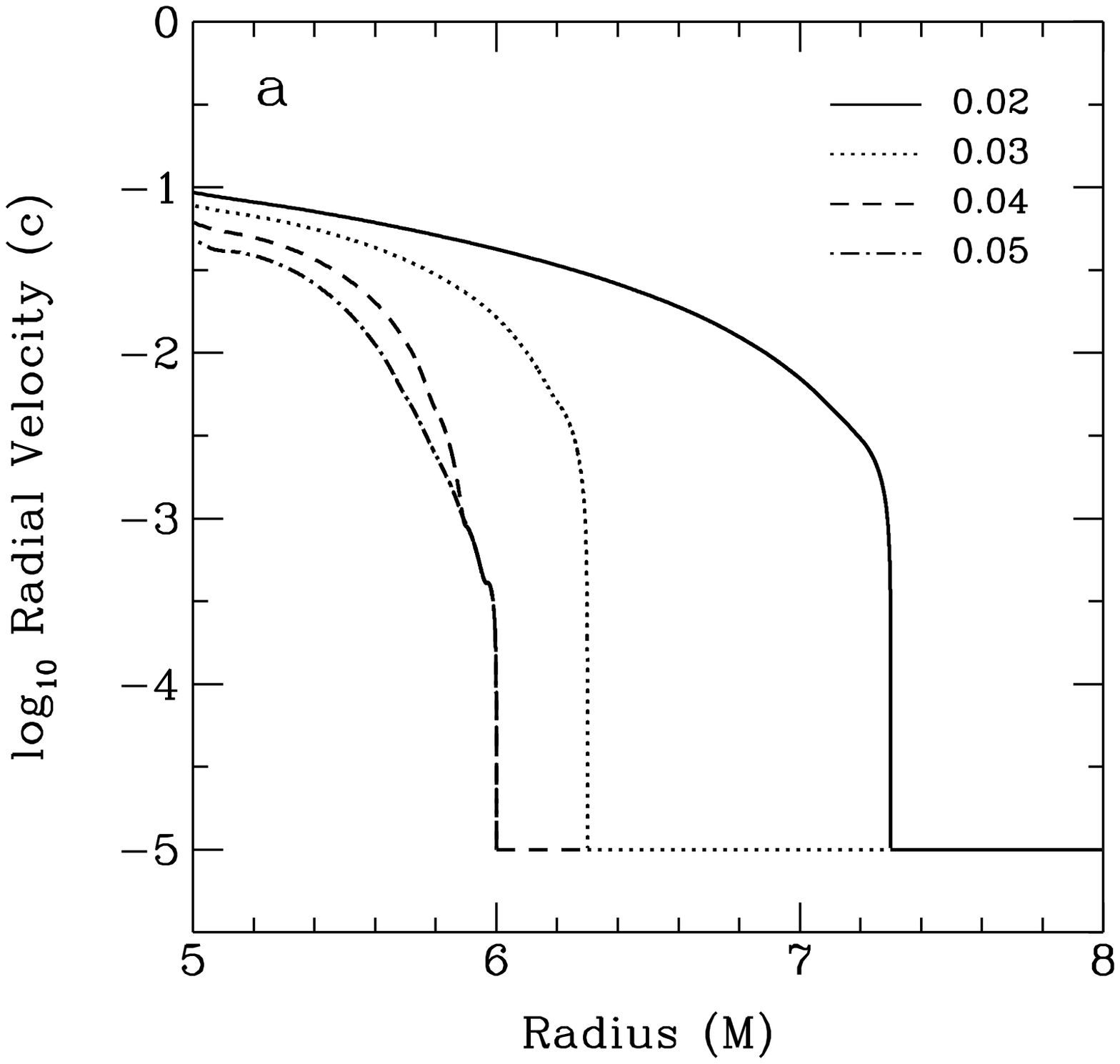,height=3.8truein,width=3.8truein}}
\hglue-1.5truecm
{\psfig{file=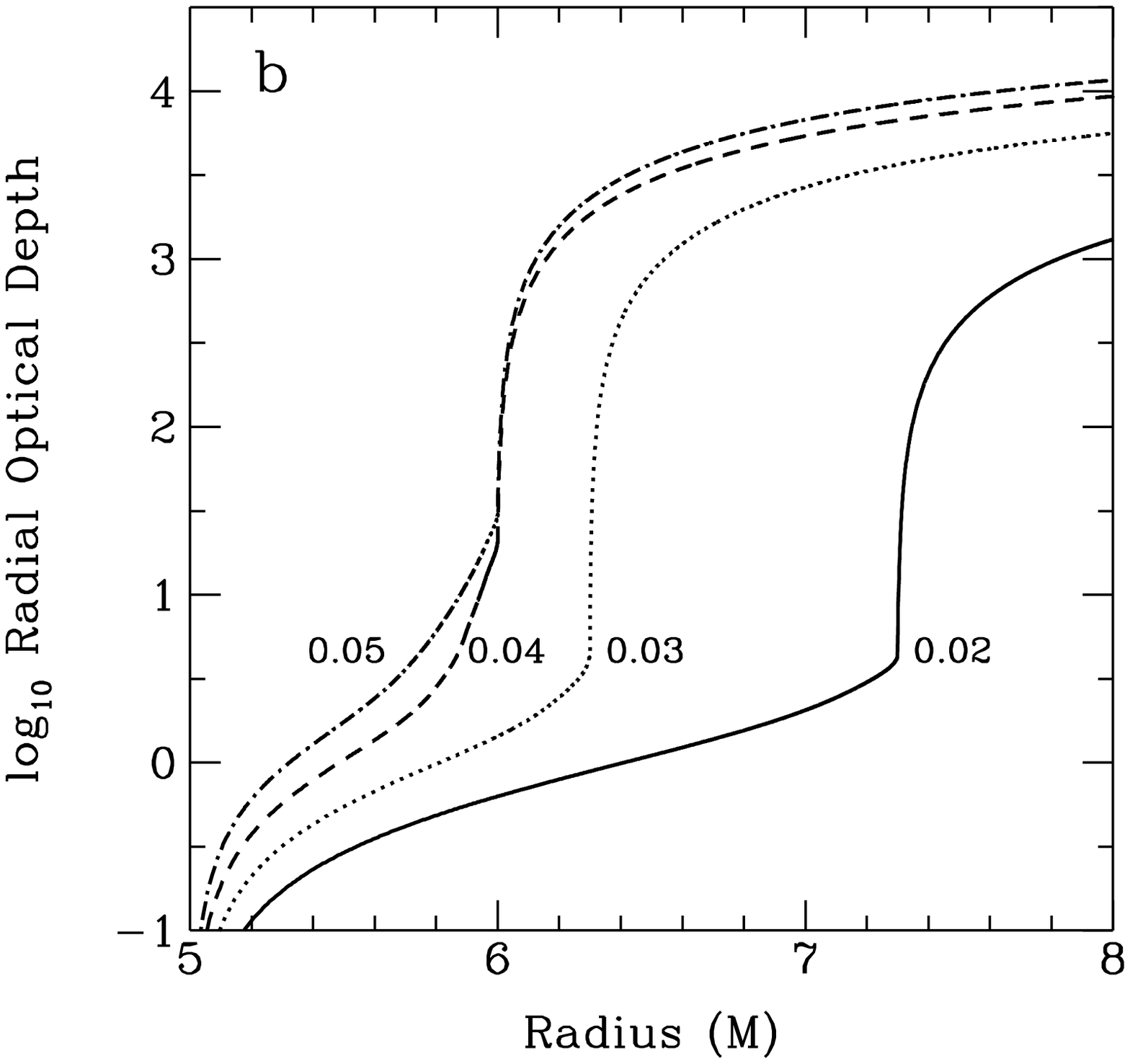,height=3.8truein,width=3.8truein}}}
 \vglue-1.5truecm
 \caption{
 Results of fully general relativistic
numerical computations \cite{MLP97} of the
gas dynamics and radiation transport in the
inner disk in the sonic-point model, for a
neutron star of radius $5M$.
 (a)~The inward radial velocity $v^{\hat
r}$ of the gas in the disk measured by a
local static observer, as a function of the
Boyer-Lindquist radial coordinate expressed
in units of the stellar mass $M$.
 (b)~The radial optical depth $\tau_r$ from
the stellar surface through the disk flow
to the radius shown on the horizontal axis.
 The four curves in each panel are labeled
with the assumed accretion rate $\mdot$
measured in units of the accretion rate
\mdote\ that would produce an accretion
luminosity at infinity equal to the
Eddington critical luminosity.
 } \label{transition} \end{figure*}

\section{THE SONIC-POINT MODEL}

Figure~\ref{spirals} shows how the QPO at
the sonic-point Keplerian frequency is
generated in the sonic-point model. Gas in
the accretion disk drifts slowly inward
until it becomes exposed to radiation from
the stellar surface. Once exposed, the gas
loses its angular momentum to radiation
drag in a radial distance $\Delta r \lta
0.01\,r$ and falls inward supersonically
along spiral trajectories like that shown
in Figure~\ref{spirals}a. Gas falling
inward from the sonic radius along spiral
trajectories collides with the neutron star
around its equator, producing an X-ray
emitting equatorial ring, which is
indicated by the grey ring around the star.

Magnetoturbulence, differential cooling, and
radiation forces create density
inhomogeneities (``clumps'') in the gas in
the accretion disk. Gas streaming inward
from clumps orbiting near the sonic radius
along trajectories with the shape shown in
Figure~\ref{spirals}a generates the density
pattern shown in Figure~\ref{spirals}b.
Collision with the stellar surface of the
denser gas from the clumps creates beams of
brighter X-ray emission, like the beam
indicated by the white dashed lines in
Figure~\ref{spirals}b, which move around the
star's equator with a frequency equal to the
orbital frequency of the clumps, generating
a quasi-periodic brightness oscillation
with frequency $\nu_{\rm Ks}$.

The QPO at the sonic-point beat frequency
is generated by weak X-ray beams that
rotate {\em with the star\/} and modulate
the radiation drag acting on the gas at the
sonic radius once each beat period. As a
result, the inward mass flux from each clump
oscillates, causing the luminosity of the
footprints to oscillate at the sonic-point
beat frequency $\nu_{\rm Bs}$.

The nature of the transition to supersonic
radial inflow is illustrated by the radial
velocity and optical depth profiles shown in
Figure~\ref{transition}. These profiles were
calculated assuming that the azimuthal
velocity of the gas in the disk is nearly
Keplerian far from the star. Internal shear
stress in the disk was assumed to create a
constant inward radial velocity $v^{\hat
r}$, as measured in the local static frame,
of $10^{-5}$. The half-height $h(r)$ of the
disk flow at radius $r$ was assumed to be
$\epsilon r$ at all radii, where $\epsilon$
is a constant and $r$ is the radius, and
the kinetic energy of the gas that collides
with the surface of the star was assumed to
be converted to radiation and emitted from
a band around the star's equator with a
half-height equal to $\epsilon R$. For
simplicity, and to show the effects of
radiation forces more clearly, any effect
of the stellar magnetic field on the gas
dynamics near the sonic transition was
neglected.

Once the drag force exerted by the radiation
from the stellar surface begins to remove
angular momentum from the gas in the
Keplerian disk, centrifugal support is lost
and the gas falls inward, accelerating
rapidly. Radiation that comes from near the
star and is scattered by the gas in the disk
is usually scattered out of the disk plane
and hence does not interact further with the
gas in the disk. Moreover, second and
successive scatterings do not contribute
proportionately to the azimuthal radiation
drag force on the gas because the radiation
field is aberrated by the first scattering
and afterward carries angular momentum
\cite{ML93,ML96}. The interaction of the
radiation with the gas in the disk was
therefore treated by assuming that the
intensity of the radiation coming from the
star is attenuated as it passes through the
gas in the disk, diminishing as
$\exp(-\tau_r)$, where $\tau_r(r)$ is the
Thomson scattering optical depth radially
outward from the stellar surface to radius
$r$, and that scattered radiation does not
contribute to removal of angular momentum
from the gas in the inner disk. In order to
simply calculation of the radiation drag
force, the differential scattering cross
section was assumed to be isotropic in the
frame comoving with the accreting gas (see
\cite{LM95}). The radiation field and the
motion of the gas were computed in full
general relativity.

\begin{figure}[t] 
 \vglue-1.25truecm
\centerline{\hglue1truecm\psfig{file=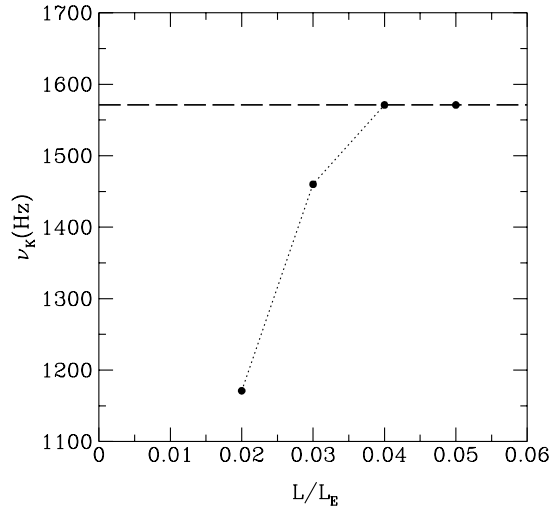,height=8.7truecm,width=8.7truecm}}
 \vglue-1.5truecm
 \caption{
 Sonic-point Keplerian frequency versus
accretion luminosity for the accretion flow
profiles shown in Figure~\ref{transition}.
From [28].
 } \label{nuVsL} \end{figure}

Figure~\ref{transition} shows that in this
model, the sonic radius decreases with
increasing accretion rate until it reaches
the radius of the ISCO, after which it stops
decreasing. Hence, the frequency of the
sonic-point Keplerian frequency QPO
increases steeply with increasing
luminosity until the sonic radius reaches
the radius of the ISCO, after which it
stops increasing, as shown in
Figure~\ref{nuVsL}.

\begin{figure}[t] 
 \vglue-0.75truecm
\centerline{\hglue0.75truecm\psfig{file=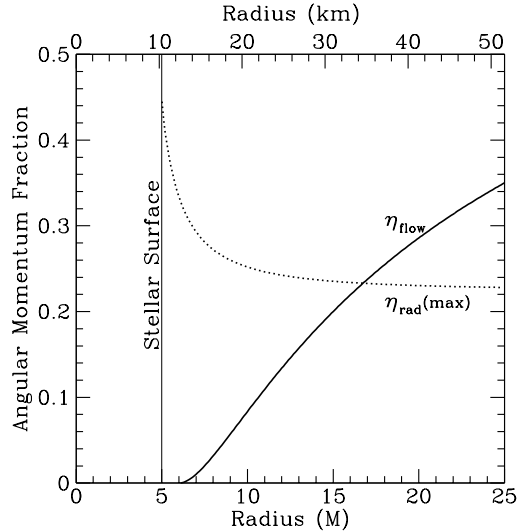,height=8.7truecm,width=8.7truecm}}
 \vglue-1.9truecm
 \caption{
 Comparison of $\eta_{\rm flow}$, the
fraction of the specific angular momentum
of an element of gas in Keplerian circular
orbit at Boyer-Lindquist radial coordinate
$r$ (measured in units of the stellar mass)
that must be removed in order for the gas
to fall from $r$ to the radius $R_{\rm ms}$
of the innermost stable circular orbit,
with $\eta_{\rm rad}({\rm max})$ the
largest fraction of the specific angular
momentum of an element of gas that can be 
removed by radiation coming from the
surface of a nonrotating, isotropically
radiating, spherical star of radius $5M$.
 } \label{AngMom} \end{figure}

The sonic point must occur between the
radius $R_{\rm ms}$ of the innermost stable
circular orbit and the radius where the two
curves shown in Figure~\ref{AngMom}
intersect, which is $\sim 3R_{\rm ms}$.
The frequency of the sonic-point Keplerian
frequency QPO is therefore restricted to a
relatively small interval for all sources.
For example, if the neutron stars in LMXBs
all have masses $\sim$1.7\,\msun, then the
frequencies of their sonic-point Keplerian
frequency QPOs would all be confined to the
range $\sim$400--1300~Hz.

\section{DISCUSSION}

The discovery with \rxte\/ of $\sim$1000~Hz
brightness oscillations from a large number
of accreting neutron stars is a spectacular
achievement that validates both the
scientific expectations that led to the
mission and the long years of hard work
that were needed to bring it to fruition.

As explained in \S3, the observations made
to date strongly favor beat-frequency
models of the kilohertz QPOs detected in
the persistent emission. The magnetospheric
beat-frequency interpretation of the
kilohertz QPO pairs can account for the fact
that their separation frequencies are
approximately constant and related to the
frequencies of the brightness oscillations
seen during X-ray bursts, but this
interpretation suffers from many serious
difficulties. The sonic-point
beat-frequency model also explains this
fact and is the most fully developed and
successful model, but it is not yet
confirmed. Therefore, it is important to
consider which further observations and
calculations would be particularly helpful
in testing the beat-frequency hypothesis and
discriminating between these two
beat-frequency models.

The detection in the persistent emission of
stable oscillations with frequencies
harmonically related to those seen during
X-ray bursts would confirm the
spin-frequency interpretation of the latter
and strongly support the beat-frequency
interpretation of the kilohertz QPO pairs.
On the other hand, detection of stable
oscillations with frequencies not
harmonically related to those seen during
X-ray bursts would seriously undermine both
models.

Beat-frequency models also predict that
weak oscillations will be present at other
special frequencies, such as the first
overtone of the beat frequency and the sum
of the spin frequency and the orbital
frequency, and detection of any of these
would provide strong support for these
models. It is therefore important to search
for QPOs at these frequencies. Even if none
are detected, upper limits on their
amplitudes would be valuable, because they
would constrain models of the coronae
around these neutron stars (see
Fig.~\ref{nsLMXBs}).

The sonic-point model predicts that the
stronger the stellar magnetic field, the
weaker the kilohertz QPOs will be
\cite{MLP97}. In the magnetospheric
interpretation, one expects just the
opposite. Figure~\ref{qpoAmps} shows that
the current data on kilohertz QPO amplitudes
clearly favors the sonic-point model, but
more precise and uniform amplitude
measurements, as well as further progress
in modeling the X-ray spectra of the Z and
atoll sources, would help to clarify the
situation.

Besides their intrinsic interest, the
$\sim$1000~Hz oscillations discovered with
\rxte\ provide a new tool with which to
probe the nature of strong gravitational
fields and the properties of dense matter.
Modeling of general relativistic effects on
the gas dynamics and radiation transport
processes involved in the generation of
kilohertz QPOs may provide evidence for the
existence of innermost stable circular
orbits \cite{MLP97,MLP98} and frame-dragging
\cite{ML93,ML96,Stella98,SV97}, both of
which are important predictions of
strong-field general relativity.

If the frequencies of the highest-frequency
kilohertz QPOs are orbital frequencies, as
in beat-frequency models, these QPOs
provide important new constraints on the
masses and radii of the neutron stars in
LMXBs and on the equation of state of
neutron star matter
\cite{KFC97,MLP97,MLP98,ZSS97}. The
brightness oscillations observed during
X-ray bursts may constrain the compactness
of neutron stars
\cite{ML98,Stroh97,Swank98}.

New \rxte\ observations are continuing to
yield significant fresh insights, and the
rapid pace of important new discoveries is
therefore likely to continue for many years.

\vskip6pt
This work was supported in part by NSF
grants AST~93-15133 and AST~96-18524 and
NASA grant NAG~5-2925 at the University of
Illinois, and NASA grant NAG~5-2868 at the
University of Chicago.


\begin{thebibliography}{9}

\vskip4pt

\bibitem{AS85} Alpar, A., \& Shaham, J.
1985, Nature, 316, 239


\bibitem{Arons92} Arons, J. 1992, ApJ, 388,
561

\bibitem{AL76a} Arons, J., \& Lea, S.\,M.
1976a, ApJ, 207, 914


%
%
%
%

\bibitem{Bradt98} Bradt, H.\,V. 1998, these
proceedings

\bibitem{BS89} Bradt, H.\,V., \& Swank
J.\,H. 1989, in Timing Neutron Stars, ed. H.
\"Ogelman \& E.P.J. van den Heuvel
(Dordrecht: Kluwer), 393

%
%
%
%
%


\bibitem{EL76} Elsner, R.\,F., \& Lamb,
F.\,K. 1976, Nature, 262, 356

\bibitem{ELP86} Epstein, R., Lamb, F.\,K.,
\& Priedhorsky, W. 1986, Astrophysics of
Time Variability in X-Ray and Gamma-Ray
Sources, Los Alamos Science, No.~13

%
%

\bibitem{Ford97b} Ford, E., et al. 1997,
ApJ, 486, L47

%

\bibitem{FLM89} Fortner, B.\,I., Lamb,
F.\,K., \& Miller, G.\,S. 1989, Nature 342,
775


\bibitem{GL79a} Ghosh, P., \& Lamb, F.\,K.
1979, ApJ, 232, 259

%

\bibitem{GL92} ---------. 1992, in X-Ray
Binaries and Recycled Pulsars, ed.
E.\,P.\,J. van den Heuvel \& S.\,A.
Rappaport (Dordrecht: Kluwer), p.~487

%
\bibitem{HK89} Hasinger, G., \& van der Klis, M.
1989, A\&A 225, 79

%

\bibitem{KFC97} Kaaret, P., Ford, E.\,C.,
\& Chen, K. 1997, ApJ, 480, L27

\bibitem{Klein96a} Klein, R.\,I., Arons, J.,
Jernigan, J.\,G., \& Hsu, J. 1996, ApJ,
457, L85

\bibitem{Klein96b} Klein, R.\,I., Jernigan,
J.\,G., Arons, J., Morgan, E.\,H., \&
Zhang, W. 1996, ApJ, 469, L119

\bibitem{Klein97} Klein, R.\,I. 1997, talk
presented at the 1997 HEAD Meeting, Estes
Park, Colorado

%
%
%
%


\bibitem{L75b} Lamb, F.~K. 1975, in Proc.
7th Texas Symposium on Relativistic
Astrophysics (Ann. NY Acad. Sci., 262, 331)



\bibitem{L89}  ---------. 1989, in Proc.\
23rd ESLAB Symp. on X-ray Astronomy, ed.\
N.\,E. White (ESA SP-296), 215

\bibitem{L91}   ---------. 1991, in Neutron
Stars: Theory and Observation, ed. J.
Ventura \& D. Pines, (Dordrecht: Kluwer),
445


\bibitem{LM95} Lamb, F.\,K., \& Miller,
M.\,C. 1995, ApJ, 439, 828

\bibitem{LP79} Lamb, F.\,K., \& Pines, D.
1979, Compact Galactic X-Ray Sources
(Urbana: Univ. of Illinois Physics Dept.)

\vskip8pt

\bibitem{L85} Lamb, F.~K., Shibazaki, N.,
Alpar, A., \& Shaham, J. 1985, Nature, 317,
681



\bibitem{Mend97b} M\'endez, M., et al.\
1997, ApJ, in press (astro-ph/9712085)

\bibitem{MP86} Middleditch, J., \&
Priedhorsky, W. 1986, ApJ, 306, 230


\bibitem{ML93} Miller, M.\,C., \& Lamb,
F.\,K. 1993, ApJ, 413, L43

\bibitem{ML96} ---------. 1996, ApJ, 470,
1033

\bibitem{ML98} ---------. 1998, ApJ,
submitted (astro- ph/9711325)

\bibitem{MLP97} Miller, M.\,C., Lamb,
F.\,K., \& Psaltis, D. 1997, ApJ, in press

\bibitem{MLP98}  ---------. 1998, these
proceedings

%
%
%
%
%
%
%
%
%

\bibitem{PL97} Psaltis, D., \& Lamb, F.\,K.
1997, in prep.

\bibitem{PL98}  ---------. 1998, in Neutron
Stars and Pulsars, ed. N. Shibazaki \& N.
Kawai (Tokyo: Universal Academy Press), in
press

\bibitem{PLM95} Psaltis, D., Lamb, F.\,K.,
\& Miller, G.\,S. 1995, ApJ, 454, L137

%
%
%
%
%
%
%
%

\bibitem{Stella98} Stella, L. 1998, these
proceedings

\bibitem{SV97} Stella, L., \& Vietri, M.
1997, ApJ, in press (astro-ph/9709085)

%
%

\bibitem{Stroh96c} Strohmayer, T., Zhang,
W., Swank, J.\,H., Smale, A., Titarchuk,
L., \& Day, C. 1996, ApJ, 469, L9

\bibitem{Stroh97} Strohmayer, T. 1997, talk
presented at the 1997 HEAD Meeting, Estes
Park, Colorado


\bibitem{Swank98} Swank, J. 1998, these
proceedings

\bibitem{Sw95} Swank, J., et al. 1995, in
The Lives of Neutron Stars, ed. M.\,A.
Alpar, \"U. K{\i}z{\i}lo{\v g}lu, \& J. van
Paradijs (Dordrecht: Kluwer), 525

\bibitem{TM1997} Titarchuk, L., \& Muslimov, A.
1997, A\&A, 323, L5

\bibitem{Unno79} Unno, W., Osaki, Y., Ando,
H., \& Shibahashi, H. 1979, Nonradial
Oscillations of Stars (Tokyo: Univ. of
Tokyo Press), \S18

%
%
%
%
%
%
%
%

\bibitem{vdK97b} van der Klis, M. 1997, in
The Many Faces of Neutron Stars, Proc. NATO
ASI, Lipari, Italy (Dordrecht: Kluwer), in
press (astro-ph/9710016)

\bibitem{vdK98} ---------. 1998, these
proceedings

\bibitem{vdK85} van der Klis, M., et al.
1985, Nature, 316, 225

\bibitem{vdK97a} van der Klis, M.,
Wijnands, R., Horne, K., \& Chen, W. 1997,
ApJ, 481, L97

%


%
%
%
%
%


%



\bibitem{ZSS97} Zhang, W., Strohmayer, T.,
\& Swank, J.\,H. 1997, ApJ, 482, L167


\end{thebibliography}
\end{document}